\documentclass[aps,prl,twocolumn,reprint,longbibliography]{revtex4-1}

\usepackage{amsmath,hyperref,graphicx,color,tikz}
\usepackage{ulem,soul}

\newsavebox{\graphicsbox}

\usepackage{txfonts}

\graphicspath{ {./}{figs/} }

\hypersetup{
    colorlinks=true,       
    linkcolor=blue,          
    citecolor=blue,        
    filecolor=blue,      
    urlcolor=blue           
}

\newcommand{\noD}{n_\mathrm{1D}}
\newcommand{\noDb}{n_\mathrm{1D}}
\newcommand{\wn}{\ell}
\newcommand{\Rm}{R}

\newcommand{\Ilab}{I_{\text{bulk}}}
\newcommand{\Irot}{I_{\text{WL}}}

\newcommand{\figwidth}{3.25in}

\newcommand{\bq}{\begin{equation}}
\newcommand{\eq}{\end{equation}}
\newcommand{\bn}{\begin{eqnarray}}
\newcommand{\en}{\end{eqnarray}}

\begin{document}

\title{Interferometric measurement of the current-phase relationship of a superfluid weak link}
\author{S. Eckel}
\email{stephen.eckel@nist.gov}
\author{F. Jendrzejewski}
\author{A. Kumar}
\author{C.J. Lobb}
\author{G.K. Campbell}
\affiliation{Joint Quantum Institute, National Institute of Standards and Technology and University of Maryland, Gaithersburg, Maryland 20899, USA}

\begin{abstract}
Weak connections between superconductors or superfluids can differ from classical links due to quantum coherence, which allows flow without resistance. Transport properties through such weak links can be described with a single function, the current-phase relationship, which serves as the quantum analog of the current-voltage relationship.  Here, we present a technique for inteferometrically measuring the current-phase relationship of superfluid weak links.  We interferometrically measure the phase gradient around a ring-shaped superfluid Bose-Einstein condensate (BEC) containing a rotating weak link, allowing us to identify the current flowing around the ring.  While our BEC weak link operates in the hydrodynamic regime, this technique can be extended to all types of weak links (including tunnel junctions) in any phase-coherent quantum gas.  Moreover, it can also measure the current-phase relationships of excitations. Such measurements may open new avenues of research in quantum transport.
\end{abstract}

\maketitle

A variety of quantum phenomena, such as Josephson effects~\cite{Josephson1964} and quantum interference~\cite{Jaklevic1964,Sato2012a}, can be observed by weakly connecting two superconductors or superfluids.  Such a weak connection can be, for example, a narrow channel or a potential barrier that allows for quantum tunneling.  For any weak link, there is a relationship between the current and the phase difference between the two superconductors or superfluids.  This current-phase relationship is essential for understanding quantum transport through the weak link~\cite{Likharex1979}. 
In superconductors, the current-phase relationship of weak links is measured routinely, and such measurements can indicate the presence of exotic quantum states, such as Majorana fermions~\cite{Fu2009,Sochnikov2013} or oscillations in the order parameter~\cite{Frolov2004}.  In superfluid liquid helium, this current-phase relationship has been measured, but only indirectly~\cite{Hoskinson2006}.  In degenerate atomic gases, the current-phase relationship has not yet been measured (although many of the effects associated with weak links, e.g., Josephson effects~\cite{Albiez2005,Levy2007}, have been observed).  Here, we interferometrically measure the phase around a ring-shaped superfluid Bose-Einstein condensate (BEC).  We use this technique to determine both the magnitude and sign of persistent currents in the ring.  In the presence of a rotating constriction that acts as a weak link, we show how to measure its current-phase relationship.

In a superfluid, the velocity $v$ is related to the gradient of the phase $\phi$ of the macroscopic wavefunction by $v=(\hbar/m)\nabla\phi$, where $\hbar$ is Planck's constant divided by $2\pi$ and $m$ is the mass of an atom.  Ignoring the transverse degrees of freedom, the number current is then $I = (\hbar \noDb/m)\nabla\phi$, where $\noD$ is the equivalent 1D density of the fluid along the direction of flow.  In a weak link, the superfluid density will vary as a function of position and velocity~\footnote{A non-linear current-phase relationship arises only if $\noD$ in the weak link depends on the velocity in the weak link.}, resulting in a potentially complicated current-phase relationship~\cite{Watanabe2009}.  For example, in an idealized Josephson junction, which is typically realized with a tunnel barrier, the phase drop across the weak link $\gamma$ is related to the current through $I = I_c \sin\gamma$, where $I_c$ is its critical current.  Because the ideal Josephson junction can be hard to achieve, the current-phase relationships of experimentally realizable weak links can exhibit higher order harmonics or become multivalued~\cite{Baratoff1096,Piazza2010,DeaverJr1972}.

In the present case, we generate a constriction that acts as a weak link in that its critical velocity is much less than that of the rest of the system~\cite{Likharex1979}.  However, our weak link is large compared to healing length of the BEC, leading to a linear current-phase relationship similar to that of a bulk superfluid.  Previous works~\cite{Albiez2005,Levy2007,Ryu2013} used weak links that operated in the tunneling regime; however, none of these measured the current-phase relationship.

Weak links have enabled manipulation of ring-shaped BECs, by both controlling a persistent current~\cite{Ramanathan2011,Moulder2012,Wright2013,Ryu2014,Eckel2014} and inducing flow between reservoirs~\cite{Ryu2013,Jendrzejewski2014}.   Because the wavefunction must be single valued, the integral of $\nabla\phi$ around any closed path must be a multiple of $2\pi$.  In particular, for a ring with mean radius $\Rm$, this leads to the constraint $(m/\hbar)\oint v(\theta)\ \Rm d\theta=2\pi \wn$, where $\theta$ is the azimuthal angle and the integer $\wn$ is a topological invariant known as the winding number.  Transitions between these quantized states can occur when a weak link stirs the superfluid at a critical rotation rate~\cite{Wright2013,Eckel2014}.  In previous experiments with ring-shaped condensates, the detection method used could only measure the magnitude of resulting winding number $\wn$.  Here, we use an interference technique to measure the phase and therefore the current flow around a ring-shaped BEC.  We demonstrate that when the rotating weak link is present, there is already a current around the ring even if $\wn=0$.  This implies that while the winding number is quantized, neither the average current nor the total angular momentum of the BEC are quantized.

\begin{figure}
	\center
	\includegraphics[width=\figwidth]{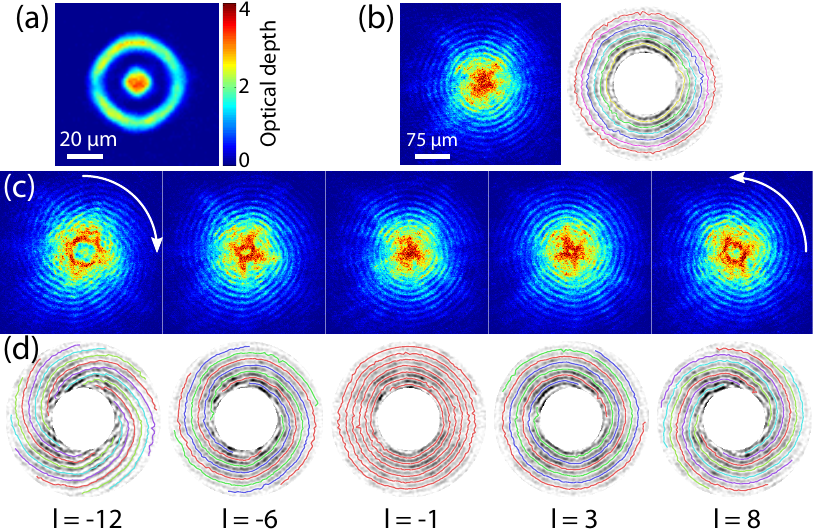}
	\caption{\label{fig:ex_winding_numbers} (a) {\it In-situ} image of the ring and disk BECs with dimensions shown. (b) Example interferogram after 15~ms time-of-flight (left) when there is no current in the ring, including traces of the azimuthal interference fringes to guide the eye (right).  (c) Interferograms for various winding numbers, where the arrow indicates the direction of flow.  (d) Traces of the interference fringes to guide the eye and count the number of spiral arms.  The extracted winding number is shown below the traces.}
\end{figure}

To measure the phase around the ring, we use two BECs of $^{23}$Na atoms held in an optical dipole trap,  as shown in Fig~\ref{fig:ex_winding_numbers}(a).  One is shaped like a disk and serves as a phase reference. The other is a concentric ring, which can sustain a persistent current.  To detect the phase of the wavefunction and thus the current in the condensate, we interfere the two separate condensates, which can be accomplished after time-of-flight (TOF) expansion. In fact, such interference experiments provided the first conclusive proof that a BEC is a single, phase-coherent object~\cite{Andrews1997}.  Later experiments used similar interference techniques to detect quantized vortices~\cite{Inouye2001}, to investigate the coherence properties of a superfluid Fermi gas~\cite{Kohstall2011}, and to study the physics of both two dimensional~\cite{Hadzibabic2006} and one-dimensional Bose gases~\cite{Hofferberth2007}.  A method similar to that presented here has been independently developed to investigate the supercurrent generated by a rapid quench through the BEC transition~\cite{Corman2014}.

Measuring the interference of our BECs after TOF expansion yields a measurement of $\psi_D^*\psi_D + \psi_D^*\psi_R + \psi_R^*\psi_D + \psi_R^*\psi_R$, where $\psi_D$ is the wavefunction of the disk and $\psi_R$ is the wavefunction of the ring.  The first term $P_D = \psi_D^*\psi_D$ produces no fringes as the disk expands.  The terms that are of most interest here contain the ring and the disk, $P_{RD} = \psi_D^*\psi_R + \psi_R^*\psi_D$, and they interfere once $\psi_R$ and $\psi_D$ expand such that they overlap.    The last term, $P_R = \psi_R^*\psi_R$, can also produce an interference pattern once the ring has expanded further, such that its characteristic width $|\sigma(t)|$ becomes comparable to $\Rm$.  At this point, the opposite sides of the ring can interfere with each other.

For simplicity, let us first consider the interference pattern when there is no weak link present and both BECs are at rest before being released from the trap [Fig.~\ref{fig:ex_winding_numbers}(b)].  Without flow, the phase is independent of angle in both the disk and ring.  The interference term $P_{RD}$ results in concentric circles.   The radial position of these azimuthal interference fringes depends on the relative phases between the two condensates; the radial separation between fringes corresponds to a phase difference of $2\pi$.  The interference term $P_R = \psi_R^*\psi_R$ produces similar concentric circles, but with a contrast that is below our detection threshold~\cite{Murray2013,Ryu2014}.

If there is no weak link present but there is a non-zero winding number in the ring, the resulting interference patterns are modified.  In this case, the phase of the ring wavefunction will be given by $\phi = \wn \theta$, assuming the ring is sufficiently smooth that both $\noD$ and $v$ are independent of the azimuthal angle $\theta$.  Such a phase profile represents a quantized persistent current:  the current takes on discrete values $\wn I_0$, where $I_0 = \noD \Omega_0 R$ and $\Omega_0 = \hbar/m\Rm^2$.  As shown in Refs.~\cite{Murray2013,Ryu2014}, the interference $P_R$ is modified in this case: a hole with quantized size appears at long times.  Previous experiments~\cite{Wright2013,Beattie2013,Eckel2014} demonstrated quantized persistent currents in a ring by releasing the BEC from a ring-shaped trap (without another BEC present) and observing the size of the resulting hole.  While this method determines the magnitude of the current, it does not determine the direction.

In addition to modifying the $P_R$ term, a persistent current also modifies the interference term $P_{RD}$, turning the $\wn=0$ concentric circles into spirals when $\wn \neq 0$ [Fig.~\ref{fig:ex_winding_numbers}(c)].  (The circular structures observed at the center of the clouds for large winding numbers are associated with the emergence of the quantized hole described by $P_R$.)  The combination of the initial azimuthal velocity of the ring atoms and the expansion of the clouds creates spirals in the interference pattern.  One can use such spirals to measure the accumulated phase around the ring $\alpha$ by tracking a maximum (or a minimum) of an interference fringe from $\theta=0$ to $\theta=2\pi$.   The net radial fringe displacement divided by the spacing between fringes yields $\alpha/2\pi$.  Because $\alpha=2\pi \wn$ in the present case, this procedure is equivalent to counting the number of spiral arms, which determines the magnitude of $\wn$, and noting their chirality, which determines its sign.

\begin{figure}
	\center
	\includegraphics[width=\figwidth]{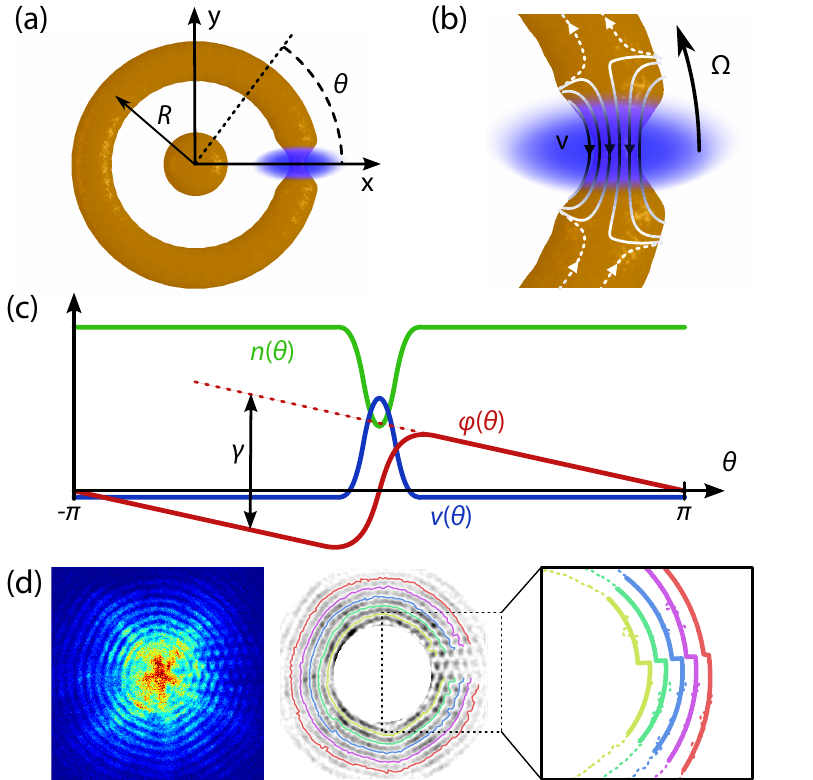}
	\caption{\label{fig:phys_fig} (a) Schematic of the atoms in the trap with a weak link applied.  The coordinate system used throughout is shown; $\theta=0$ corresponds to the $\hat{x}$ axis.  (b) A close up of the weak link region.  When the weak link is rotated at $\Omega$, atoms flow through the weak link as shown by the stream lines.  Larger velocities along the stream lines correspond to darker lines.  (c) The resulting density $n(\theta)$, velocity $v(\theta)$, and phase $\phi(\theta)$ as a function of angle, with the phase drop $\gamma$ across the weak link shown.  (d) Method of extracting the the phase from an interferogram  (left).  First, we trace the interference fringes around the ring (center) and then fit the discontinuity across the region where the barrier was (right).}
\end{figure}

Adding the weak link modifies the interference pattern beyond the spirals described above.  The weak link, as shown in Fig.~\ref{fig:phys_fig}(a)--(b), is a density-depleted region in the ring.  Once the cloud is released, atoms from either side of the density-depleted weak link expand toward each other and interfere, causing additional interference fringes to appear in the radial direction, as shown in Fig.~\ref{fig:phys_fig}(d).  Just as in the case where there was no weak link, we can still measure $\alpha$ by tracking the azimuthal interference fringes around the ring, excluding the weak link region.  To measure their radial displacement after going from $\theta=0$ to $\theta=2\pi$, one must extrapolate those fringes back through the weak link region.  Dividing the size of the extrapolated radial displacement of a single fringe by the spacing between the fringes once again yields $\alpha/2\pi$.  Here, $\alpha$ is not necessarily a multiple of $2\pi$.  This measurement of $\alpha$ allows us to extract the current-phase relationship of the weak link, as shown below.

Before discussing the results, we first describe the experimental techniques.  The ring and the disk traps are formed by the combination of two crossed lasers.  A red-detuned laser shaped like a sheet creates vertical confinement, while an intensity-masked blue-detuned laser separates the ring trap from the disk trap to form the two BECs.  A blue-detuned laser generates the weak link by creating a Gaussian-shaped repulsive potential of height $U$ and $1/e^2$ full-width of $\approx 6$~$\mu$m (for details on the weak link, see Ref.~\cite{Jendrzejewski2014}).  This potential depletes the density in a small portion of the ring, as shown in Fig.~\ref{fig:phys_fig}(a).   On average, a total of $\approx8\times10^5$ atoms reside in the traps.  The ring BEC has a mean radius of $22.4(4)$~$\mu$m and annular width (twice the Thomas-Fermi radius) of $\approx6$~$\mu$m.  It contains $\approx75$~\% of the atoms and has an initial chemical potential $\mu_0/\hbar \approx 2\pi\times(3$~kHz$)$.  The central disk contains $\approx 25$~\% of the atoms and has a Thomas-Fermi radius of $\approx 5$~$\mu$m.  While the disk is approximately hard-walled, the ring is closer to harmonic with a measured radial trapping frequency of $\approx390$~Hz.  The distance between the inner radius of the ring and the disk is $\approx 6.5$~$\mu$m.

To prepare the system in a well defined quantized persistent current state with a chosen $\wn$, we stir our weak link at a corresponding $\Omega$.  Such stirring lasts for 1~s, during which the rotation rate of the weak link is constant but the strength of the weak link potential ramps on linearly in 300~ms, holds constant for 400~ms, and ramps off in another 300~ms.  To measure the resulting $\wn$, we hold the BECs for an additional 100~ms, then release them, and lastly, image the interference pattern after 15~ms TOF expansion.  This procedure produced the data shown in Fig.~\ref{fig:ex_winding_numbers}(b)--(c).

We extend these results by measuring $\alpha$ in the presence of a weak link as a function of $U$, the rotation rate $\Omega$, and the initial winding number $\wn$.   First, we stir to set the initial winding number $\wn=0$ or $\pm1$, as described above.  To get the highest fidelity for setting $\wn$ (95(2)~\%), 
we empirically find that $U\approx1.2\mu_0$ and $\Omega\approx\pm0.9$~Hz or zero, depending on which winding number state we wish to initialize.  Second, we stir for 1~s at a new $\Omega$ and $U$.  During the first 300~ms of this second stage of stirring, the weak link potential ramps from zero to the chosen $U$, and afterwards remains constant.  We adjust the starting position such that the weak link is at $\theta=0$ at the end of this second step.  At this point, the trap and weak link potential turn off, which releases the cloud.  After 17~ms TOF (for slightly better resolution), we again image the cloud.

\begin{figure}
 \center
 \includegraphics[width=\figwidth]{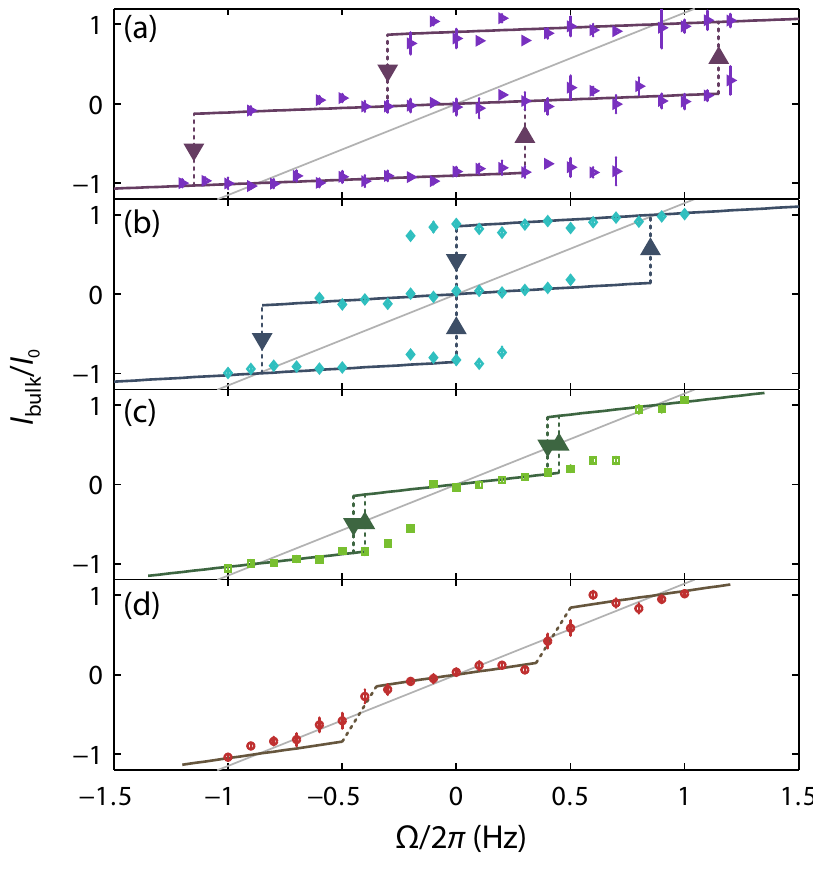}
 \caption{\label{fig:total_flux} Plot of the normalized current around the bulk of the ring, $\Ilab/I_0=\alpha/2\pi$, vs. the rotation rate $\Omega$ of the weak link for four different weak link potential stengths $U$: (a) $0.45\ \mu_0$, (b) $0.6\ \mu_0$, (c) $0.7\ \mu_0$, (d) $0.8\ \mu_0$.  The solid lines are the prediction of our model (see text).  The dashed, vertical lines show the predicted transitions between the different winding number branches. The thin, gray, diagonal lines represent the case where all the atoms move around the bulk of the ring with the weak link, i.e., $\Ilab=\noDb R \Omega$.}
\end{figure}

The above procedures result in a measurement of the phase accumulated around the ring, $\alpha$, which is related to the current around the bulk of the ring through $\Ilab=\noD(m/\hbar)\nabla\phi = \noD(m/\hbar)(\alpha/2\pi R)$.  We measure $\Ilab$, normalized to $I_0=\noDb \Omega_0 \Rm$, as a function of $\Omega$ for a variety of different $U$; Fig.~\ref{fig:total_flux} shows four examples.    As shown, there are discrete jumps in $\Ilab$ at specific rotations rates.  At these critical rotation rates, the system experiences a phase slip which changes $\wn$.   These critical rotation rates are dependent on $U$, and can be hysteretic.   Fig.~\ref{fig:total_flux}(a)--(b) show such hysteresis.  The size of the hysteresis loop is consistent with previous measurements~\cite{Eckel2014}.   In addition, we measure a non-zero, superfluid $\Ilab$ for rotation rates below the critical rotation rate, where presumably there are no excitations and $\wn=0$.

$\Ilab$ can be understood in the following way:  
As the weak link rotates around the ring, it must displace superfluid from in front of its path and superfluid must fill in behind it.  The number of atoms that must flow per unit time is proportional to the difference in the density in the weak link and the bulk of the ring.  If the flow was only confined to the weak link and in the direction opposite of the rotation, as shown by the solid stream lines in Fig.~\ref{fig:phys_fig}(b), it would violate the condition $\oint v(\theta)\ Rd\theta=0$.  Thus, the atoms in the bulk of the ring must have some velocity in the same direction as the rotation (dashed flow lines) in order to cancel the phase accumulated by the atoms moving through the weak link, as shown in Fig.~\ref{fig:phys_fig}(c).  This is analogous to fluxoid vs. flux quantization in superconductors~\cite{TilleyandTilley}: although $\wn$ must always be quantized, neither the current $\Ilab$ or the total angular momentum are (see Supplemental).

\begin{figure}
 \center
 \includegraphics[width=\figwidth]{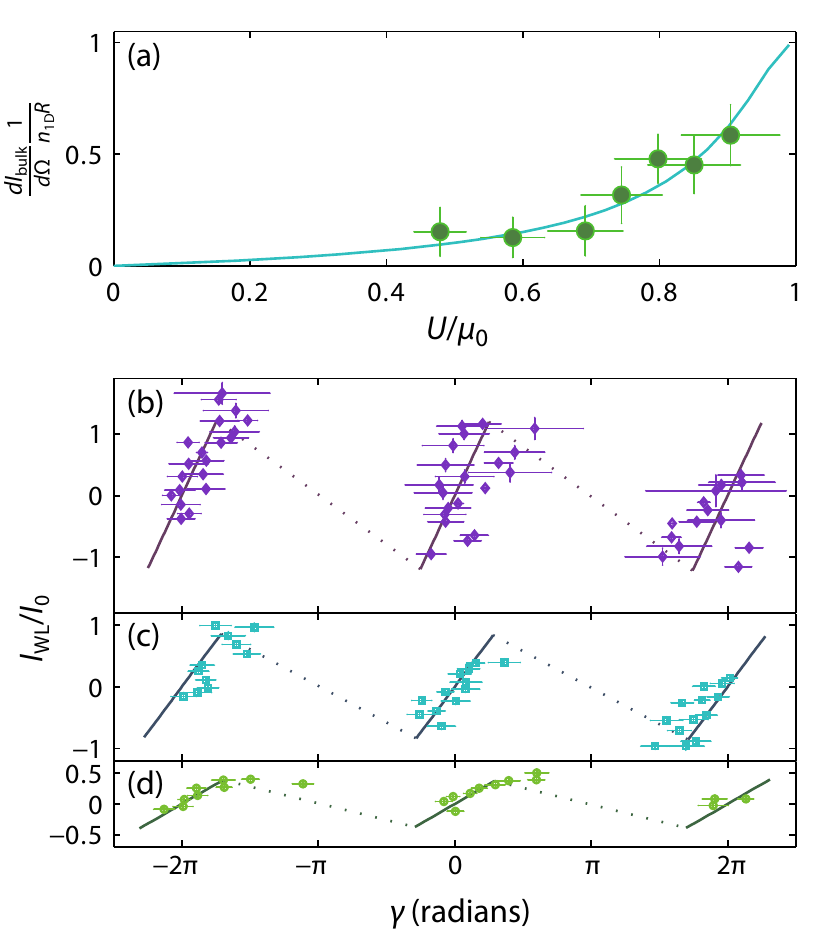}
 \caption{\label{fig:well_known_effect} (a) Derivative of the initial bulk current $d\Ilab/d\Omega$ vs. $U$, normalized to the expected value in the limit where $U/\mu_0\geq1$, $\noD R$.  The solid line shows the prediction of the LDA model. (b)--(d) Extracted current-phase relationships from the data in Fig.~\ref{fig:total_flux}, for three different weak link potential strengths $U$: (b) $0.45\ \mu_0$, (c) $0.6\ \mu_0$, (d) $0.7\ \mu_0$.   $\gamma$ is the phase across the weak link and $\Irot$ is the current through it, normalized to $I_0=\noD R\Omega_0\approx5\times10^5$~atoms/s.  The solid curves represent the prediction of our theoretical model.  The dashed lines merely guide the eye by connecting the multiple branches of the current-phase relationship.}
\end{figure}

Initially, $\Ilab$ vs. $\Omega$ is linear, and Fig.~\ref{fig:well_known_effect}(a) shows its measured derivative $d\Ilab/d\Omega|_{\Omega=0}$ as a function of $U$.  As $U\rightarrow0$, no atoms move, and $\Ilab=0$ for all $\Omega$.  For a given rotation, increasing $U$ displaces more atoms, resulting in a larger current around the bulk of the ring.  As expected, $d\Ilab/d\Omega|_{\Omega=0}$ continues to increase until $U=\mu_0$, at which point no atoms can move through the weak link and they all must move around the ring, i.e., $\Ilab = \noD R\Omega$.  This limit corresponds to solid-body rotation; Fig.~\ref{fig:total_flux} shows this limit as thin gray lines.  In a reference frame that rotates with the weak link, there is no flow in this limit and thus $\Irot=0$, where $\Irot$ is the current in the weak link's frame.  The opposite limit of $U\rightarrow 0$ corresponds to $\Irot = \noD R\Omega$ (where we have taken $\Irot>0$ to represent flow that is opposite the rotation).

The $\Ilab/I_0$ vs. $\Omega$ curves of Fig.~\ref{fig:total_flux} can be predicted using a model based on the local density approximation (LDA) to the Gross-Pitaevksii equation (see Supplemental and Ref~\cite{Watanabe2009}), but assuming a critical velocity as measured in Ref.~\cite{Eckel2014}.  (An LDA treatment can be used because the azimuthal length of the weak link of $\approx6$~$\mu$m is larger than the healing length $\xi=\sqrt{\hbar^2/2 m \mu_0} \approx0.3$~$\mu$m.)  All parameters are measured independently; none are adjustable.   The predictions of this model are shown in Figs.~\ref{fig:total_flux} and~\ref{fig:well_known_effect}(a) as the solid curves.

The current-phase relationship is best evaluated in the weak link's frame, where $\Irot = \noD R\Omega - \Ilab$.  The phase drop across the weak link, $\gamma$, that corresponds to $\Irot$ is given by $\gamma = -2\pi(\Ilab/I_0)$ [see Fig.~\ref{fig:phys_fig}(c)].  For a constant $\Omega$, the current-phase relationship determines how much current flows past the weak link ($\Irot$, measured in the weak link's frame) and how much flows past a fixed point in the bulk of the ring ($\Ilab$).  Using these relationships, we can extract the current phase-relationship from the data in Fig.~\ref{fig:total_flux}, the results of which are shown in Fig.~\ref{fig:well_known_effect}(b)--(d).

For our BEC system, our model predicts that the current-phase relationship is roughly linear.  Non-linearities caused by changes in the superfluid density with $\gamma$ occur when the velocity through the weak link nears the speed of sound; however, because our critical velocity is lower than the speed of sound, these non-linearities are small.  Thus, our weak link is far from an ideal Josephson junction.  We also note that our simple model cannot predict the current-phase relationship in the region indicated by the dotted line in Fig.~\ref{fig:well_known_effect}(b)--(d).  (The dotted lines merely guide the eye between the predicted branches.)  In this branch, we expect dissipation to play a key role in the dynamics.

Ideally, one would want to apply our method to a weak link that could be tuned from the hydrodynamic flow regime observed here to the Josephson or tunneling regime.  For the ideal Josephson junction with a sinusoidal current-phase relationship, the $\Ilab$ vs. $\Omega$ lines in Fig~\ref{fig:total_flux} would be curved, a signature that has yet to be observed in degenerate atomic gases.  To obtain such a signature, one would need a potential barrier whose width is comparable to the healing length of the condensate to suppress hydrodynamic flow but allow quantum mechanical tunneling.

In conclusion, we have demonstrated a technique for measuring the current-phase relationship of a weak link in a dilute-gas superfluid BEC.  We demonstrate that a rotating weak link always generates a superfluid current in the bulk of the ring, even when the rotation rate is less than any critical velocity in the system.  The magnitude of that current is determined by the current-phase relationship.  Our new method will allow for better characterization of weak links and, in the case of a tunnel junction, should provide the signature of the existence of idealized Josephson junctions in BEC systems.  In addition, measurement of the current-phase relationship enables prediction of the hysteretic energy landscape of our system~\cite{Eckel2014}, which, like the energy landscape of a flux qubit, should be quantized~\cite{Clarke2008}.  More broadly, it is possible that this method can be extended to measure the current-phase relationships of various excitations, such as solitonic-vortices~\cite{Ku2014}.  Lastly, this powerful tool may prove important for studying transport in other, exotic forms of quantum matter, such as unitary Fermi gases~\cite{Yefsah2013}, Tonks-Giradeau gases~\cite{Paredes2004, Kinoshita2004}, and quasi-2D condensates near the BKT transition~\cite{Hadzibabic2006}.

\begin{acknowledgments}
The authors thank J.G. Lee for technical assistance, M. Edwards, E. Tiesinga, R. Mathew, and W. T. Hill III for useful discussions.  We thank W.D. Phillips for an extremely thorough reading of the manuscript.  This work was partially supported by ONR, the ARO atomtronics MURI, and the NSF through the PFC at the JQI. S. Eckel is supported by a National Research Council postdoctoral fellowship.
\end{acknowledgments}


\begin{thebibliography}{0}%
\makeatletter
\providecommand \@ifxundefined [1]{%
 \@ifx{#1\undefined}
}%
\providecommand \@ifnum [1]{%
 \ifnum #1\expandafter \@firstoftwo
 \else \expandafter \@secondoftwo
 \fi
}%
\providecommand \@ifx [1]{%
 \ifx #1\expandafter \@firstoftwo
 \else \expandafter \@secondoftwo
 \fi
}%
\providecommand \natexlab [1]{#1}%
\providecommand \enquote  [1]{``#1''}%
\providecommand \bibnamefont  [1]{#1}%
\providecommand \bibfnamefont [1]{#1}%
\providecommand \citenamefont [1]{#1}%
\providecommand \href@noop [0]{\@secondoftwo}%
\providecommand \href [0]{\begingroup \@sanitize@url \@href}%
\providecommand \@href[1]{\@@startlink{#1}\@@href}%
\providecommand \@@href[1]{\endgroup#1\@@endlink}%
\providecommand \@sanitize@url [0]{\catcode `\\12\catcode `\$12\catcode
  `\&12\catcode `\#12\catcode `\^12\catcode `\_12\catcode `\%12\relax}%
\providecommand \@@startlink[1]{}%
\providecommand \@@endlink[0]{}%
\providecommand \url  [0]{\begingroup\@sanitize@url \@url }%
\providecommand \@url [1]{\endgroup\@href {#1}{\urlprefix }}%
\providecommand \urlprefix  [0]{URL }%
\providecommand \Eprint [0]{\href }%
\providecommand \doibase [0]{http://dx.doi.org/}%
\providecommand \selectlanguage [0]{\@gobble}%
\providecommand \bibinfo  [0]{\@secondoftwo}%
\providecommand \bibfield  [0]{\@secondoftwo}%
\providecommand \translation [1]{[#1]}%
\providecommand \BibitemOpen [0]{}%
\providecommand \bibitemStop [0]{}%
\providecommand \bibitemNoStop [0]{.\EOS\space}%
\providecommand \EOS [0]{\spacefactor3000\relax}%
\providecommand \BibitemShut  [1]{\csname bibitem#1\endcsname}%
\let\auto@bib@innerbib\@empty
\end{thebibliography}%


\begin{thebibliography}{36}%
\makeatletter
\providecommand \@ifxundefined [1]{%
 \@ifx{#1\undefined}
}%
\providecommand \@ifnum [1]{%
 \ifnum #1\expandafter \@firstoftwo
 \else \expandafter \@secondoftwo
 \fi
}%
\providecommand \@ifx [1]{%
 \ifx #1\expandafter \@firstoftwo
 \else \expandafter \@secondoftwo
 \fi
}%
\providecommand \natexlab [1]{#1}%
\providecommand \enquote  [1]{``#1''}%
\providecommand \bibnamefont  [1]{#1}%
\providecommand \bibfnamefont [1]{#1}%
\providecommand \citenamefont [1]{#1}%
\providecommand \href@noop [0]{\@secondoftwo}%
\providecommand \href [0]{\begingroup \@sanitize@url \@href}%
\providecommand \@href[1]{\@@startlink{#1}\@@href}%
\providecommand \@@href[1]{\endgroup#1\@@endlink}%
\providecommand \@sanitize@url [0]{\catcode `\\12\catcode `\$12\catcode
  `\&12\catcode `\#12\catcode `\^12\catcode `\_12\catcode `\%12\relax}%
\providecommand \@@startlink[1]{}%
\providecommand \@@endlink[0]{}%
\providecommand \url  [0]{\begingroup\@sanitize@url \@url }%
\providecommand \@url [1]{\endgroup\@href {#1}{\urlprefix }}%
\providecommand \urlprefix  [0]{URL }%
\providecommand \Eprint [0]{\href }%
\providecommand \doibase [0]{http://dx.doi.org/}%
\providecommand \selectlanguage [0]{\@gobble}%
\providecommand \bibinfo  [0]{\@secondoftwo}%
\providecommand \bibfield  [0]{\@secondoftwo}%
\providecommand \translation [1]{[#1]}%
\providecommand \BibitemOpen [0]{}%
\providecommand \bibitemStop [0]{}%
\providecommand \bibitemNoStop [0]{.\EOS\space}%
\providecommand \EOS [0]{\spacefactor3000\relax}%
\providecommand \BibitemShut  [1]{\csname bibitem#1\endcsname}%
\let\auto@bib@innerbib\@empty
\bibitem [{\citenamefont {Josephson}(1964)}]{Josephson1964}%
  \BibitemOpen
  \bibfield  {author} {\bibinfo {author} {\bibfnamefont {B.}~\bibnamefont
  {Josephson}},\ }\bibfield  {title} {\enquote {\bibinfo {title} {{Coupled
  Superconductors}},}\ }\href {\doibase 10.1103/RevModPhys.36.216} {\bibfield
  {journal} {\bibinfo  {journal} {Rev. Mod. Phys.}\ }\textbf {\bibinfo {volume}
  {36}},\ \bibinfo {pages} {216--220} (\bibinfo {year} {1964})}\BibitemShut
  {NoStop}%
\bibitem [{\citenamefont {Jaklevic}\ \emph {et~al.}(1964)\citenamefont
  {Jaklevic}, \citenamefont {Lambe}, \citenamefont {Silver},\ and\
  \citenamefont {Mercereau}}]{Jaklevic1964}%
  \BibitemOpen
  \bibfield  {author} {\bibinfo {author} {\bibfnamefont {RC}~\bibnamefont
  {Jaklevic}}, \bibinfo {author} {\bibfnamefont {John}\ \bibnamefont {Lambe}},
  \bibinfo {author} {\bibfnamefont {AH}~\bibnamefont {Silver}}, \ and\ \bibinfo
  {author} {\bibfnamefont {JE}~\bibnamefont {Mercereau}},\ }\bibfield  {title}
  {\enquote {\bibinfo {title} {{Quantum Interference Effects in Josephson
  Tunneling}},}\ }\href {\doibase 10.1103/PhysRevLett.12.159} {\bibfield
  {journal} {\bibinfo  {journal} {Phys. Rev. Lett.}\ }\textbf {\bibinfo
  {volume} {12}},\ \bibinfo {pages} {159--160} (\bibinfo {year}
  {1964})}\BibitemShut {NoStop}%
\bibitem [{\citenamefont {Sato}\ and\ \citenamefont
  {Packard}(2012)}]{Sato2012a}%
  \BibitemOpen
  \bibfield  {author} {\bibinfo {author} {\bibfnamefont {Y}~\bibnamefont
  {Sato}}\ and\ \bibinfo {author} {\bibfnamefont {R~E}\ \bibnamefont
  {Packard}},\ }\bibfield  {title} {\enquote {\bibinfo {title} {{Superfluid
  helium quantum interference devices: physics and applications.}}}\ }\href
  {\doibase 10.1088/0034-4885/75/1/016401} {\bibfield  {journal} {\bibinfo
  {journal} {Rep. Prog. Phys.}\ }\textbf {\bibinfo {volume} {75}},\ \bibinfo
  {pages} {016401} (\bibinfo {year} {2012})}\BibitemShut {NoStop}%
\bibitem [{\citenamefont {Likharev}(1979)}]{Likharex1979}%
  \BibitemOpen
  \bibfield  {author} {\bibinfo {author} {\bibfnamefont {K.K.}\ \bibnamefont
  {Likharev}},\ }\bibfield  {title} {\enquote {\bibinfo {title}
  {{Superconducting weak links}},}\ }\href {\doibase 10.1103/RevModPhys.51.101}
  {\bibfield  {journal} {\bibinfo  {journal} {Rev. Mod. Phys.}\ }\textbf
  {\bibinfo {volume} {51}},\ \bibinfo {pages} {101--159} (\bibinfo {year}
  {1979})}\BibitemShut {NoStop}%
\bibitem [{\citenamefont {Fu}\ and\ \citenamefont {Kane}(2009)}]{Fu2009}%
  \BibitemOpen
  \bibfield  {author} {\bibinfo {author} {\bibfnamefont {Liang}\ \bibnamefont
  {Fu}}\ and\ \bibinfo {author} {\bibfnamefont {C.}~\bibnamefont {Kane}},\
  }\bibfield  {title} {\enquote {\bibinfo {title} {{Josephson current and noise
  at a superconductor/quantum-spin-Hall-insulator/superconductor junction}},}\
  }\href {\doibase 10.1103/PhysRevB.79.161408} {\bibfield  {journal} {\bibinfo
  {journal} {Phys. Rev. B}\ }\textbf {\bibinfo {volume} {79}},\ \bibinfo
  {pages} {161408} (\bibinfo {year} {2009})}\BibitemShut {NoStop}%
\bibitem [{\citenamefont {Sochnikov}\ \emph {et~al.}(2013)\citenamefont
  {Sochnikov}, \citenamefont {Bestwick}, \citenamefont {Williams},
  \citenamefont {Lippman}, \citenamefont {Fisher}, \citenamefont
  {Goldhaber-Gordon}, \citenamefont {Kirtley},\ and\ \citenamefont
  {Moler}}]{Sochnikov2013}%
  \BibitemOpen
  \bibfield  {author} {\bibinfo {author} {\bibfnamefont {Ilya}\ \bibnamefont
  {Sochnikov}}, \bibinfo {author} {\bibfnamefont {Andrew~J}\ \bibnamefont
  {Bestwick}}, \bibinfo {author} {\bibfnamefont {James~R}\ \bibnamefont
  {Williams}}, \bibinfo {author} {\bibfnamefont {Thomas~M}\ \bibnamefont
  {Lippman}}, \bibinfo {author} {\bibfnamefont {Ian~R}\ \bibnamefont {Fisher}},
  \bibinfo {author} {\bibfnamefont {David}\ \bibnamefont {Goldhaber-Gordon}},
  \bibinfo {author} {\bibfnamefont {John~R}\ \bibnamefont {Kirtley}}, \ and\
  \bibinfo {author} {\bibfnamefont {Kathryn~A}\ \bibnamefont {Moler}},\
  }\bibfield  {title} {\enquote {\bibinfo {title} {{Direct measurement of
  current-phase relations in superconductor/topological
  insulator/superconductor junctions.}}}\ }\href {\doibase 10.1021/nl400997k}
  {\bibfield  {journal} {\bibinfo  {journal} {Nano Lett.}\ }\textbf {\bibinfo
  {volume} {13}},\ \bibinfo {pages} {3086--92} (\bibinfo {year}
  {2013})}\BibitemShut {NoStop}%
\bibitem [{\citenamefont {Frolov}\ \emph {et~al.}(2004)\citenamefont {Frolov},
  \citenamefont {{Van Harlingen}}, \citenamefont {Oboznov}, \citenamefont
  {Bolginov},\ and\ \citenamefont {Ryazanov}}]{Frolov2004}%
  \BibitemOpen
  \bibfield  {author} {\bibinfo {author} {\bibfnamefont {S.}~\bibnamefont
  {Frolov}}, \bibinfo {author} {\bibfnamefont {D.}~\bibnamefont {{Van
  Harlingen}}}, \bibinfo {author} {\bibfnamefont {V.}~\bibnamefont {Oboznov}},
  \bibinfo {author} {\bibfnamefont {V.}~\bibnamefont {Bolginov}}, \ and\
  \bibinfo {author} {\bibfnamefont {V.}~\bibnamefont {Ryazanov}},\ }\bibfield
  {title} {\enquote {\bibinfo {title} {{Measurement of the current-phase
  relation of superconductor/ferromagnet/superconductor $\pi$ Josephson
  junctions}},}\ }\href {\doibase 10.1103/PhysRevB.70.144505} {\bibfield
  {journal} {\bibinfo  {journal} {Phys. Rev. B}\ }\textbf {\bibinfo {volume}
  {70}},\ \bibinfo {pages} {144505} (\bibinfo {year} {2004})}\BibitemShut
  {NoStop}%
\bibitem [{\citenamefont {Hoskinson}\ \emph {et~al.}(2005)\citenamefont
  {Hoskinson}, \citenamefont {Sato}, \citenamefont {Hahn},\ and\ \citenamefont
  {Packard}}]{Hoskinson2006}%
  \BibitemOpen
  \bibfield  {author} {\bibinfo {author} {\bibfnamefont {E}~\bibnamefont
  {Hoskinson}}, \bibinfo {author} {\bibfnamefont {Y}~\bibnamefont {Sato}},
  \bibinfo {author} {\bibfnamefont {I}~\bibnamefont {Hahn}}, \ and\ \bibinfo
  {author} {\bibfnamefont {R~E}\ \bibnamefont {Packard}},\ }\bibfield  {title}
  {\enquote {\bibinfo {title} {{Transition from phase slips to the Josephson
  effect in a superfluid $^4$He weak link}},}\ }\href {\doibase 10.1038/nphys190}
  {\bibfield  {journal} {\bibinfo  {journal} {Nat. Phys.}\ }\textbf {\bibinfo
  {volume} {2}},\ \bibinfo {pages} {23--26} (\bibinfo {year}
  {2005})}\BibitemShut {NoStop}%
\bibitem [{\citenamefont {Albiez}\ \emph {et~al.}(2005)\citenamefont {Albiez},
  \citenamefont {Gati}, \citenamefont {F\"{o}lling}, \citenamefont {Hunsmann},
  \citenamefont {Cristiani},\ and\ \citenamefont {Oberthaler}}]{Albiez2005}%
  \BibitemOpen
  \bibfield  {author} {\bibinfo {author} {\bibfnamefont {Michael}\ \bibnamefont
  {Albiez}}, \bibinfo {author} {\bibfnamefont {Rudolf}\ \bibnamefont {Gati}},
  \bibinfo {author} {\bibfnamefont {Jonas}\ \bibnamefont {F\"{o}lling}},
  \bibinfo {author} {\bibfnamefont {Stefan}\ \bibnamefont {Hunsmann}}, \bibinfo
  {author} {\bibfnamefont {Matteo}\ \bibnamefont {Cristiani}}, \ and\ \bibinfo
  {author} {\bibfnamefont {Markus~K.}\ \bibnamefont {Oberthaler}},\ }\bibfield
  {title} {\enquote {\bibinfo {title} {{Direct Observation of Tunneling and
  Nonlinear Self-Trapping in a Single Bosonic Josephson Junction}},}\ }\href
  {\doibase 10.1103/PhysRevLett.95.010402} {\bibfield  {journal} {\bibinfo
  {journal} {Phys. Rev. Lett.}\ }\textbf {\bibinfo {volume} {95}},\ \bibinfo
  {pages} {010402} (\bibinfo {year} {2005})}\BibitemShut {NoStop}%
\bibitem [{\citenamefont {Levy}\ \emph {et~al.}(2007)\citenamefont {Levy},
  \citenamefont {Lahoud}, \citenamefont {Shomroni},\ and\ \citenamefont
  {Steinhauer}}]{Levy2007}%
  \BibitemOpen
  \bibfield  {author} {\bibinfo {author} {\bibfnamefont {S}~\bibnamefont
  {Levy}}, \bibinfo {author} {\bibfnamefont {E}~\bibnamefont {Lahoud}},
  \bibinfo {author} {\bibfnamefont {I}~\bibnamefont {Shomroni}}, \ and\
  \bibinfo {author} {\bibfnamefont {J}~\bibnamefont {Steinhauer}},\ }\bibfield
  {title} {\enquote {\bibinfo {title} {{The a.c. and d.c. Josephson effects in
  a Bose-Einstein condensate}},}\ }\href {\doibase 10.1038/nature06186}
  {\bibfield  {journal} {\bibinfo  {journal} {Nature}\ }\textbf {\bibinfo
  {volume} {449}},\ \bibinfo {pages} {579--583} (\bibinfo {year}
  {2007})}\BibitemShut {NoStop}%
\bibitem [{Note1()}]{Note1}%
  \BibitemOpen
  \bibinfo {note} {A non-linear current-phase relationship arises only if
  $n_\protect \mathrm {1D}$ in the weak link depends on the velocity in the
  weak link.}\BibitemShut {Stop}%
\bibitem [{\citenamefont {Watanabe}\ \emph {et~al.}(2009)\citenamefont
  {Watanabe}, \citenamefont {Dalfovo}, \citenamefont {Piazza}, \citenamefont
  {Pitaevskii},\ and\ \citenamefont {Stringari}}]{Watanabe2009}%
  \BibitemOpen
  \bibfield  {author} {\bibinfo {author} {\bibfnamefont {Gentaro}\ \bibnamefont
  {Watanabe}}, \bibinfo {author} {\bibfnamefont {F}~\bibnamefont {Dalfovo}},
  \bibinfo {author} {\bibfnamefont {F}~\bibnamefont {Piazza}}, \bibinfo
  {author} {\bibfnamefont {L.}~\bibnamefont {Pitaevskii}}, \ and\ \bibinfo
  {author} {\bibfnamefont {S}~\bibnamefont {Stringari}},\ }\bibfield  {title}
  {\enquote {\bibinfo {title} {{Critical velocity of superfluid flow through
  single-barrier and periodic potentials}},}\ }\href {\doibase
  10.1103/PhysRevA.80.053602} {\bibfield  {journal} {\bibinfo  {journal} {Phys.
  Rev. A}\ }\textbf {\bibinfo {volume} {80}},\ \bibinfo {pages} {53602}
  (\bibinfo {year} {2009})}\BibitemShut {NoStop}%
\bibitem [{\citenamefont {Baratoff}\ \emph {et~al.}(1970)\citenamefont
  {Baratoff}, \citenamefont {Blackburn},\ and\ \citenamefont
  {Schwartz}}]{Baratoff1096}%
  \BibitemOpen
  \bibfield  {author} {\bibinfo {author} {\bibfnamefont {Alexis}\ \bibnamefont
  {Baratoff}}, \bibinfo {author} {\bibfnamefont {James~A}\ \bibnamefont
  {Blackburn}}, \ and\ \bibinfo {author} {\bibfnamefont {Brian~B}\ \bibnamefont
  {Schwartz}},\ }\bibfield  {title} {\enquote {\bibinfo {title} {{Current-Phase
  Relationship in Short Superconducting Weak Leaks}},}\ }\href {\doibase
  10.1103/PhysRevLett.25.1096} {\bibfield  {journal} {\bibinfo  {journal}
  {Phys. Rev. Lett.}\ }\textbf {\bibinfo {volume} {25}},\ \bibinfo {pages}
  {1096} (\bibinfo {year} {1970})}\BibitemShut {NoStop}%
\bibitem [{\citenamefont {Piazza}\ \emph {et~al.}(2010)\citenamefont {Piazza},
  \citenamefont {Collins},\ and\ \citenamefont {Smerzi}}]{Piazza2010}%
  \BibitemOpen
  \bibfield  {author} {\bibinfo {author} {\bibfnamefont {F}~\bibnamefont
  {Piazza}}, \bibinfo {author} {\bibfnamefont {L~A}\ \bibnamefont {Collins}}, \
  and\ \bibinfo {author} {\bibfnamefont {A}~\bibnamefont {Smerzi}},\ }\bibfield
   {title} {\enquote {\bibinfo {title} {{Current-phase relation of a
  Bose-Einstein condensate flowing through a weak link}},}\ }\href {\doibase
  10.1103/PhysRevA.81.033613} {\bibfield  {journal} {\bibinfo  {journal} {Phys.
  Rev. A}\ }\textbf {\bibinfo {volume} {81}},\ \bibinfo {pages} {33613}
  (\bibinfo {year} {2010})}\BibitemShut {NoStop}%
\bibitem [{\citenamefont {{Deaver Jr.}}\ and\ \citenamefont
  {Pierce}(1972)}]{DeaverJr1972}%
  \BibitemOpen
  \bibfield  {author} {\bibinfo {author} {\bibfnamefont {B~S}\ \bibnamefont
  {{Deaver Jr.}}}\ and\ \bibinfo {author} {\bibfnamefont {J~M}\ \bibnamefont
  {Pierce}},\ }\bibfield  {title} {\enquote {\bibinfo {title} {{Relaxation
  oscillator model for superconducting bridges}},}\ }\href {\doibase
  10.1016/0375-9601(72)90496-3} {\bibfield  {journal} {\bibinfo  {journal}
  {Phys. Lett. A}\ }\textbf {\bibinfo {volume} {38}},\ \bibinfo {pages}
  {81--82} (\bibinfo {year} {1972})}\BibitemShut {NoStop}%
\bibitem [{\citenamefont {Ryu}\ \emph {et~al.}(2013)\citenamefont {Ryu},
  \citenamefont {Blackburn}, \citenamefont {Blinova},\ and\ \citenamefont
  {Boshier}}]{Ryu2013}%
  \BibitemOpen
  \bibfield  {author} {\bibinfo {author} {\bibfnamefont {C.}~\bibnamefont
  {Ryu}}, \bibinfo {author} {\bibfnamefont {P.~W.}\ \bibnamefont {Blackburn}},
  \bibinfo {author} {\bibfnamefont {a.~a.}\ \bibnamefont {Blinova}}, \ and\
  \bibinfo {author} {\bibfnamefont {M.~G.}\ \bibnamefont {Boshier}},\
  }\bibfield  {title} {\enquote {\bibinfo {title} {{Experimental Realization of
  Josephson Junctions for an Atom SQUID}},}\ }\href {\doibase
  10.1103/PhysRevLett.111.205301} {\bibfield  {journal} {\bibinfo  {journal}
  {Phys. Rev. Lett.}\ }\textbf {\bibinfo {volume} {111}},\ \bibinfo {pages}
  {205301} (\bibinfo {year} {2013})}\BibitemShut {NoStop}%
\bibitem [{\citenamefont {Ramanathan}\ \emph {et~al.}(2011)\citenamefont
  {Ramanathan}, \citenamefont {Wright}, \citenamefont {Muniz}, \citenamefont
  {Zelan}, \citenamefont {Hill}, \citenamefont {Lobb}, \citenamefont
  {Helmerson}, \citenamefont {Phillips},\ and\ \citenamefont
  {Campbell}}]{Ramanathan2011}%
  \BibitemOpen
  \bibfield  {author} {\bibinfo {author} {\bibfnamefont {A.}~\bibnamefont
  {Ramanathan}}, \bibinfo {author} {\bibfnamefont {K.~C.}\ \bibnamefont
  {Wright}}, \bibinfo {author} {\bibfnamefont {S.~R.}\ \bibnamefont {Muniz}},
  \bibinfo {author} {\bibfnamefont {M.}~\bibnamefont {Zelan}}, \bibinfo
  {author} {\bibfnamefont {W.~T.}\ \bibnamefont {Hill}}, \bibinfo {author}
  {\bibfnamefont {C.~J.}\ \bibnamefont {Lobb}}, \bibinfo {author}
  {\bibfnamefont {K.}~\bibnamefont {Helmerson}}, \bibinfo {author}
  {\bibfnamefont {W.~D.}\ \bibnamefont {Phillips}}, \ and\ \bibinfo {author}
  {\bibfnamefont {G.~K.}\ \bibnamefont {Campbell}},\ }\bibfield  {title}
  {\enquote {\bibinfo {title} {{Superflow in a Toroidal Bose-Einstein
  Condensate: An Atom Circuit with a Tunable Weak Link}},}\ }\href {\doibase
  10.1103/PhysRevLett.106.130401} {\bibfield  {journal} {\bibinfo  {journal}
  {Phys. Rev. Lett.}\ }\textbf {\bibinfo {volume} {106}},\ \bibinfo {pages}
  {130401} (\bibinfo {year} {2011})}\BibitemShut {NoStop}%
\bibitem [{\citenamefont {Moulder}\ \emph {et~al.}(2012)\citenamefont
  {Moulder}, \citenamefont {Beattie}, \citenamefont {Smith}, \citenamefont
  {Tammuz},\ and\ \citenamefont {Hadzibabic}}]{Moulder2012}%
  \BibitemOpen
  \bibfield  {author} {\bibinfo {author} {\bibfnamefont {Stuart}\ \bibnamefont
  {Moulder}}, \bibinfo {author} {\bibfnamefont {Scott}\ \bibnamefont
  {Beattie}}, \bibinfo {author} {\bibfnamefont {Robert~P.}\ \bibnamefont
  {Smith}}, \bibinfo {author} {\bibfnamefont {Naaman}\ \bibnamefont {Tammuz}},
  \ and\ \bibinfo {author} {\bibfnamefont {Zoran}\ \bibnamefont {Hadzibabic}},\
  }\bibfield  {title} {\enquote {\bibinfo {title} {{Quantized supercurrent
  decay in an annular Bose-Einstein condensate}},}\ }\href {\doibase
  10.1103/PhysRevA.86.013629} {\bibfield  {journal} {\bibinfo  {journal} {Phys.
  Rev. A}\ }\textbf {\bibinfo {volume} {86}},\ \bibinfo {pages} {013629}
  (\bibinfo {year} {2012})}\BibitemShut {NoStop}%
\bibitem [{\citenamefont {Wright}\ \emph {et~al.}(2013)\citenamefont {Wright},
  \citenamefont {Blakestad}, \citenamefont {Lobb}, \citenamefont {Phillips},\
  and\ \citenamefont {Campbell}}]{Wright2013}%
  \BibitemOpen
  \bibfield  {author} {\bibinfo {author} {\bibfnamefont {K.}~\bibnamefont
  {Wright}}, \bibinfo {author} {\bibfnamefont {R.}~\bibnamefont {Blakestad}},
  \bibinfo {author} {\bibfnamefont {C.}~\bibnamefont {Lobb}}, \bibinfo {author}
  {\bibfnamefont {W.}~\bibnamefont {Phillips}}, \ and\ \bibinfo {author}
  {\bibfnamefont {G.}~\bibnamefont {Campbell}},\ }\bibfield  {title} {\enquote
  {\bibinfo {title} {{Driving Phase Slips in a Superfluid Atom Circuit with a
  Rotating Weak Link}},}\ }\href {\doibase 10.1103/PhysRevLett.110.025302}
  {\bibfield  {journal} {\bibinfo  {journal} {Phys. Rev. Lett.}\ }\textbf
  {\bibinfo {volume} {110}},\ \bibinfo {pages} {25302} (\bibinfo {year}
  {2013})}\BibitemShut {NoStop}%
\bibitem [{\citenamefont {Ryu}\ \emph {et~al.}(2014)\citenamefont {Ryu},
  \citenamefont {Henderson},\ and\ \citenamefont {Boshier}}]{Ryu2014}%
  \BibitemOpen
  \bibfield  {author} {\bibinfo {author} {\bibfnamefont {C}~\bibnamefont
  {Ryu}}, \bibinfo {author} {\bibfnamefont {K~C}\ \bibnamefont {Henderson}}, \
  and\ \bibinfo {author} {\bibfnamefont {M~G}\ \bibnamefont {Boshier}},\
  }\bibfield  {title} {\enquote {\bibinfo {title} {{Creation of matter wave
  Bessel beams and observation of quantized circulation in a Bose–Einstein
  condensate}},}\ }\href {\doibase 10.1088/1367-2630/16/1/013046} {\bibfield
  {journal} {\bibinfo  {journal} {New J. Phys.}\ }\textbf {\bibinfo {volume}
  {16}},\ \bibinfo {pages} {013046} (\bibinfo {year} {2014})}\BibitemShut
  {NoStop}%
\bibitem [{\citenamefont {Eckel}\ \emph {et~al.}(2014)\citenamefont {Eckel},
  \citenamefont {Lee}, \citenamefont {Jendrzejewski}, \citenamefont {Murray},
  \citenamefont {Clark}, \citenamefont {Lobb}, \citenamefont {Phillips},
  \citenamefont {Edwards},\ and\ \citenamefont {Campbell}}]{Eckel2014}%
  \BibitemOpen
  \bibfield  {author} {\bibinfo {author} {\bibfnamefont {Stephen}\ \bibnamefont
  {Eckel}}, \bibinfo {author} {\bibfnamefont {Jeffrey~G.}\ \bibnamefont {Lee}},
  \bibinfo {author} {\bibfnamefont {Fred}\ \bibnamefont {Jendrzejewski}},
  \bibinfo {author} {\bibfnamefont {Noel}\ \bibnamefont {Murray}}, \bibinfo
  {author} {\bibfnamefont {Charles~W.}\ \bibnamefont {Clark}}, \bibinfo
  {author} {\bibfnamefont {Christopher~J.}\ \bibnamefont {Lobb}}, \bibinfo
  {author} {\bibfnamefont {William~D.}\ \bibnamefont {Phillips}}, \bibinfo
  {author} {\bibfnamefont {Mark}\ \bibnamefont {Edwards}}, \ and\ \bibinfo
  {author} {\bibfnamefont {Gretchen~K.}\ \bibnamefont {Campbell}},\ }\bibfield
  {title} {\enquote {\bibinfo {title} {{Hysteresis in a quantized superfluid
  ‘atomtronic’ circuit}},}\ }\href {\doibase 10.1038/nature12958}
  {\bibfield  {journal} {\bibinfo  {journal} {Nature}\ }\textbf {\bibinfo
  {volume} {506}},\ \bibinfo {pages} {200--203} (\bibinfo {year}
  {2014})}\BibitemShut {NoStop}%
\bibitem [{\citenamefont {Jendrzejewski}\ \emph {et~al.}(2014)\citenamefont
  {Jendrzejewski}, \citenamefont {Eckel}, \citenamefont {Murray}, \citenamefont
  {Lanier}, \citenamefont {Edwards}, \citenamefont {Lobb},\ and\ \citenamefont
  {Campbell}}]{Jendrzejewski2014}%
  \BibitemOpen
  \bibfield  {author} {\bibinfo {author} {\bibfnamefont {F}~\bibnamefont
  {Jendrzejewski}}, \bibinfo {author} {\bibfnamefont {S}~\bibnamefont {Eckel}},
  \bibinfo {author} {\bibfnamefont {N}~\bibnamefont {Murray}}, \bibinfo
  {author} {\bibfnamefont {C}~\bibnamefont {Lanier}}, \bibinfo {author}
  {\bibfnamefont {M}~\bibnamefont {Edwards}}, \bibinfo {author} {\bibfnamefont
  {C.~J.}\ \bibnamefont {Lobb}}, \ and\ \bibinfo {author} {\bibfnamefont
  {G.~K.}\ \bibnamefont {Campbell}},\ }\bibfield  {title} {\enquote {\bibinfo
  {title} {{Resistive Flow in a Weakly Interacting Bose-Einstein
  Condensate}},}\ }\href
  {http://link.aps.org/doi/10.1103/PhysRevLett.113.045305} {\bibfield
  {journal} {\bibinfo  {journal} {Phys. Rev. Lett.}\ }\textbf {\bibinfo
  {volume} {113}},\ \bibinfo {pages} {045305} (\bibinfo {year}
  {2014})}\BibitemShut {NoStop}%
\bibitem [{\citenamefont {Andrews}\ \emph {et~al.}(1997)\citenamefont
  {Andrews}, \citenamefont {Townsend}, \citenamefont {Miesner}, \citenamefont
  {Durfree}, \citenamefont {Kurn},\ and\ \citenamefont
  {Ketterle}}]{Andrews1997}%
  \BibitemOpen
  \bibfield  {author} {\bibinfo {author} {\bibfnamefont {M.~R.}\ \bibnamefont
  {Andrews}}, \bibinfo {author} {\bibfnamefont {C.~G.}\ \bibnamefont
  {Townsend}}, \bibinfo {author} {\bibfnamefont {H.-J.}\ \bibnamefont
  {Miesner}}, \bibinfo {author} {\bibfnamefont {D.~S.}\ \bibnamefont
  {Durfree}}, \bibinfo {author} {\bibfnamefont {D.~M.}\ \bibnamefont {Kurn}}, \
  and\ \bibinfo {author} {\bibfnamefont {W.}~\bibnamefont {Ketterle}},\
  }\bibfield  {title} {\enquote {\bibinfo {title} {{Observation of Interference
  Between Two Bose Condensates}},}\ }\href {\doibase
  10.1126/science.275.5300.637} {\bibfield  {journal} {\bibinfo  {journal}
  {Science}\ }\textbf {\bibinfo {volume} {275}},\ \bibinfo {pages} {637}
  (\bibinfo {year} {1997})}\BibitemShut {NoStop}%
\bibitem [{\citenamefont {Inouye}\ \emph {et~al.}(2001)\citenamefont {Inouye},
  \citenamefont {Gupta}, \citenamefont {Rosenband}, \citenamefont {Chikkatur},
  \citenamefont {G\"{o}rlitz}, \citenamefont {Gustavson}, \citenamefont
  {Leanhardt}, \citenamefont {Pritchard},\ and\ \citenamefont
  {Ketterle}}]{Inouye2001}%
  \BibitemOpen
  \bibfield  {author} {\bibinfo {author} {\bibfnamefont {S.}~\bibnamefont
  {Inouye}}, \bibinfo {author} {\bibfnamefont {S.}~\bibnamefont {Gupta}},
  \bibinfo {author} {\bibfnamefont {T.}~\bibnamefont {Rosenband}}, \bibinfo
  {author} {\bibfnamefont {A.}~\bibnamefont {Chikkatur}}, \bibinfo {author}
  {\bibfnamefont {A.}~\bibnamefont {G\"{o}rlitz}}, \bibinfo {author}
  {\bibfnamefont {T.}~\bibnamefont {Gustavson}}, \bibinfo {author}
  {\bibfnamefont {A.}~\bibnamefont {Leanhardt}}, \bibinfo {author}
  {\bibfnamefont {D.}~\bibnamefont {Pritchard}}, \ and\ \bibinfo {author}
  {\bibfnamefont {W.}~\bibnamefont {Ketterle}},\ }\bibfield  {title} {\enquote
  {\bibinfo {title} {{Observation of Vortex Phase Singularities in
  Bose-Einstein Condensates}},}\ }\href {\doibase
  10.1103/PhysRevLett.87.080402} {\bibfield  {journal} {\bibinfo  {journal}
  {Phys. Rev. Lett.}\ }\textbf {\bibinfo {volume} {87}},\ \bibinfo {pages}
  {080402} (\bibinfo {year} {2001})}\BibitemShut {NoStop}%
\bibitem [{\citenamefont {Kohstall}\ \emph {et~al.}(2011)\citenamefont
  {Kohstall}, \citenamefont {Riedl}, \citenamefont {{S\'{a}nchez Guajardo}},
  \citenamefont {Sidorenkov}, \citenamefont {{Hecker Denschlag}},\ and\
  \citenamefont {Grimm}}]{Kohstall2011}%
  \BibitemOpen
  \bibfield  {author} {\bibinfo {author} {\bibfnamefont {C}~\bibnamefont
  {Kohstall}}, \bibinfo {author} {\bibfnamefont {S}~\bibnamefont {Riedl}},
  \bibinfo {author} {\bibfnamefont {E~R}\ \bibnamefont {{S\'{a}nchez
  Guajardo}}}, \bibinfo {author} {\bibfnamefont {L~a}\ \bibnamefont
  {Sidorenkov}}, \bibinfo {author} {\bibfnamefont {J}~\bibnamefont {{Hecker
  Denschlag}}}, \ and\ \bibinfo {author} {\bibfnamefont {R}~\bibnamefont
  {Grimm}},\ }\bibfield  {title} {\enquote {\bibinfo {title} {{Observation of
  interference between two molecular Bose–Einstein condensates}},}\ }\href
  {\doibase 10.1088/1367-2630/13/6/065027} {\bibfield  {journal} {\bibinfo
  {journal} {New J. Phys.}\ }\textbf {\bibinfo {volume} {13}},\ \bibinfo
  {pages} {065027} (\bibinfo {year} {2011})}\BibitemShut {NoStop}%
\bibitem [{\citenamefont {Hadzibabic}\ \emph {et~al.}(2006)\citenamefont
  {Hadzibabic}, \citenamefont {Kr\"{u}ger}, \citenamefont {Cheneau},
  \citenamefont {Battelier},\ and\ \citenamefont {Dalibard}}]{Hadzibabic2006}%
  \BibitemOpen
  \bibfield  {author} {\bibinfo {author} {\bibfnamefont {Zoran}\ \bibnamefont
  {Hadzibabic}}, \bibinfo {author} {\bibfnamefont {Peter}\ \bibnamefont
  {Kr\"{u}ger}}, \bibinfo {author} {\bibfnamefont {Marc}\ \bibnamefont
  {Cheneau}}, \bibinfo {author} {\bibfnamefont {Baptiste}\ \bibnamefont
  {Battelier}}, \ and\ \bibinfo {author} {\bibfnamefont {Jean}\ \bibnamefont
  {Dalibard}},\ }\bibfield  {title} {\enquote {\bibinfo {title}
  {{Berezinskii-Kosterlitz-Thouless crossover in a trapped atomic gas.}}}\
  }\href {\doibase 10.1038/nature04851} {\bibfield  {journal} {\bibinfo
  {journal} {Nature}\ }\textbf {\bibinfo {volume} {441}},\ \bibinfo {pages}
  {1118--21} (\bibinfo {year} {2006})}\BibitemShut {NoStop}%
\bibitem [{\citenamefont {Hofferberth}\ \emph {et~al.}(2007)\citenamefont
  {Hofferberth}, \citenamefont {Lesanovsky}, \citenamefont {Fischer},
  \citenamefont {Schumm},\ and\ \citenamefont
  {Schmiedmayer}}]{Hofferberth2007}%
  \BibitemOpen
  \bibfield  {author} {\bibinfo {author} {\bibfnamefont {S}~\bibnamefont
  {Hofferberth}}, \bibinfo {author} {\bibfnamefont {I}~\bibnamefont
  {Lesanovsky}}, \bibinfo {author} {\bibfnamefont {B}~\bibnamefont {Fischer}},
  \bibinfo {author} {\bibfnamefont {T}~\bibnamefont {Schumm}}, \ and\ \bibinfo
  {author} {\bibfnamefont {J}~\bibnamefont {Schmiedmayer}},\ }\bibfield
  {title} {\enquote {\bibinfo {title} {{Non-equilibrium coherence dynamics in
  one-dimensional Bose gases.}}}\ }\href {\doibase 10.1038/nature06149}
  {\bibfield  {journal} {\bibinfo  {journal} {Nature}\ }\textbf {\bibinfo
  {volume} {449}},\ \bibinfo {pages} {324--7} (\bibinfo {year}
  {2007})}\BibitemShut {NoStop}%
\bibitem [{\citenamefont {Corman}\ \emph {et~al.}(2014)\citenamefont {Corman},
  \citenamefont {Chomaz}, \citenamefont {Bienaim\'{e}}, \citenamefont
  {Desbuquois}, \citenamefont {Weintenberg}, \citenamefont {Nascimbene},
  \citenamefont {Dalibard},\ and\ \citenamefont {Beugnon}}]{Corman2014}%
  \BibitemOpen
  \bibfield  {author} {\bibinfo {author} {\bibfnamefont {Laura}\ \bibnamefont
  {Corman}}, \bibinfo {author} {\bibfnamefont {Lauriane}\ \bibnamefont
  {Chomaz}}, \bibinfo {author} {\bibfnamefont {Tom}\ \bibnamefont
  {Bienaim\'{e}}}, \bibinfo {author} {\bibfnamefont {R\'{e}mi}\ \bibnamefont
  {Desbuquois}}, \bibinfo {author} {\bibfnamefont {Christof}\ \bibnamefont
  {Weintenberg}}, \bibinfo {author} {\bibfnamefont {Sylvain}\ \bibnamefont
  {Nascimbene}}, \bibinfo {author} {\bibfnamefont {Jean}\ \bibnamefont
  {Dalibard}}, \ and\ \bibinfo {author} {\bibfnamefont {J\'{e}r\^{o}me}\
  \bibnamefont {Beugnon}},\ }\bibfield  {title} {\enquote {\bibinfo {title}
  {{Quench-induced supercurrents in an annular two-dimensional Bose gas}},}\
  }\href {http://arxiv.org/abs/1406.4073} {\ ,\ \bibinfo {pages} {1--7}
  (\bibinfo {year} {2014})},\ \Eprint {http://arxiv.org/abs/1406.4073}
  {arXiv:1406.4073} \BibitemShut {NoStop}%
\bibitem [{\citenamefont {Murray}\ \emph {et~al.}(2013)\citenamefont {Murray},
  \citenamefont {Krygier}, \citenamefont {Edwards}, \citenamefont {Wright},
  \citenamefont {Campbell},\ and\ \citenamefont {Clark}}]{Murray2013}%
  \BibitemOpen
  \bibfield  {author} {\bibinfo {author} {\bibfnamefont {Noel}\ \bibnamefont
  {Murray}}, \bibinfo {author} {\bibfnamefont {Michael}\ \bibnamefont
  {Krygier}}, \bibinfo {author} {\bibfnamefont {Mark}\ \bibnamefont {Edwards}},
  \bibinfo {author} {\bibfnamefont {K.~C.}\ \bibnamefont {Wright}}, \bibinfo
  {author} {\bibfnamefont {G.~K.}\ \bibnamefont {Campbell}}, \ and\ \bibinfo
  {author} {\bibfnamefont {Charles~W.}\ \bibnamefont {Clark}},\ }\bibfield
  {title} {\enquote {\bibinfo {title} {{Probing the circulation of ring-shaped
  Bose-Einstein condensates}},}\ }\href {\doibase 10.1103/PhysRevA.88.053615}
  {\bibfield  {journal} {\bibinfo  {journal} {Phys. Rev. A}\ }\textbf {\bibinfo
  {volume} {88}},\ \bibinfo {pages} {053615} (\bibinfo {year}
  {2013})}\BibitemShut {NoStop}%
\bibitem [{\citenamefont {Beattie}\ \emph {et~al.}(2013)\citenamefont
  {Beattie}, \citenamefont {Moulder}, \citenamefont {Fletcher},\ and\
  \citenamefont {Hadzibabic}}]{Beattie2013}%
  \BibitemOpen
  \bibfield  {author} {\bibinfo {author} {\bibfnamefont {Scott}\ \bibnamefont
  {Beattie}}, \bibinfo {author} {\bibfnamefont {Stuart}\ \bibnamefont
  {Moulder}}, \bibinfo {author} {\bibfnamefont {Richard~J.}\ \bibnamefont
  {Fletcher}}, \ and\ \bibinfo {author} {\bibfnamefont {Zoran}\ \bibnamefont
  {Hadzibabic}},\ }\bibfield  {title} {\enquote {\bibinfo {title} {{Persistent
  Currents in Spinor Condensates}},}\ }\href {\doibase
  10.1103/PhysRevLett.110.025301} {\bibfield  {journal} {\bibinfo  {journal}
  {Phys. Rev. Lett.}\ }\textbf {\bibinfo {volume} {110}},\ \bibinfo {pages}
  {025301} (\bibinfo {year} {2013})}\BibitemShut {NoStop}%
\bibitem [{\citenamefont {Tilley}\ and\ \citenamefont
  {Tilley}(1990)}]{TilleyandTilley}%
  \BibitemOpen
  \bibfield  {author} {\bibinfo {author} {\bibfnamefont {David~R}\ \bibnamefont
  {Tilley}}\ and\ \bibinfo {author} {\bibfnamefont {John}\ \bibnamefont
  {Tilley}},\ }\href@noop {} {\emph {\bibinfo {title} {{Superfluidity and
  Superconductivity}}}},\ \bibinfo {edition} {3rd}\ ed.,\ edited by\ \bibinfo
  {editor} {\bibfnamefont {Douglas~F}\ \bibnamefont {Brewer}}\ (\bibinfo
  {publisher} {Adam Hilger},\ \bibinfo {year} {1990})\BibitemShut {NoStop}%
\bibitem [{\citenamefont {Clarke}\ and\ \citenamefont
  {Wilhelm}(2008)}]{Clarke2008}%
  \BibitemOpen
  \bibfield  {author} {\bibinfo {author} {\bibfnamefont {John}\ \bibnamefont
  {Clarke}}\ and\ \bibinfo {author} {\bibfnamefont {Frank~K}\ \bibnamefont
  {Wilhelm}},\ }\bibfield  {title} {\enquote {\bibinfo {title}
  {{Superconducting quantum bits}},}\ }\href {\doibase 10.1038/nature07128}
  {\bibfield  {journal} {\bibinfo  {journal} {Nature}\ }\textbf {\bibinfo
  {volume} {453}},\ \bibinfo {pages} {1031--1043} (\bibinfo {year}
  {2008})}\BibitemShut {NoStop}%
\bibitem [{\citenamefont {Ku}\ \emph {et~al.}(2014)\citenamefont {Ku},
  \citenamefont {Ji}, \citenamefont {Mukherjee}, \citenamefont
  {Guardado-Sanchez}, \citenamefont {Cheuk}, \citenamefont {Yefsah},\ and\
  \citenamefont {Zwierlein}}]{Ku2014}%
  \BibitemOpen
  \bibfield  {author} {\bibinfo {author} {\bibfnamefont {Mark J.~H.}\
  \bibnamefont {Ku}}, \bibinfo {author} {\bibfnamefont {Wenjie}\ \bibnamefont
  {Ji}}, \bibinfo {author} {\bibfnamefont {Biswaroop}\ \bibnamefont
  {Mukherjee}}, \bibinfo {author} {\bibfnamefont {Elmer}\ \bibnamefont
  {Guardado-Sanchez}}, \bibinfo {author} {\bibfnamefont {Lawrence~W.}\
  \bibnamefont {Cheuk}}, \bibinfo {author} {\bibfnamefont {Tarik}\ \bibnamefont
  {Yefsah}}, \ and\ \bibinfo {author} {\bibfnamefont {Martin~W.}\ \bibnamefont
  {Zwierlein}},\ }\bibfield  {title} {\enquote {\bibinfo {title} {{Motion of a
  Solitonic Vortex in the BEC-BCS Crossover}},}\ }\href
  {http://link.aps.org/doi/10.1103/PhysRevLett.113.065301} {\bibfield
  {journal} {\bibinfo  {journal} {Phys. Rev. Lett.}\ }\textbf {\bibinfo
  {volume} {113}},\ \bibinfo {pages} {065301} (\bibinfo {year}
  {2014})}\BibitemShut {NoStop}%
\bibitem [{\citenamefont {Yefsah}\ \emph {et~al.}(2013)\citenamefont {Yefsah},
  \citenamefont {Sommer}, \citenamefont {Ku}, \citenamefont {Cheuk},
  \citenamefont {Ji}, \citenamefont {Bakr},\ and\ \citenamefont
  {Zwierlein}}]{Yefsah2013}%
  \BibitemOpen
  \bibfield  {author} {\bibinfo {author} {\bibfnamefont {Tarik}\ \bibnamefont
  {Yefsah}}, \bibinfo {author} {\bibfnamefont {Ariel~T}\ \bibnamefont
  {Sommer}}, \bibinfo {author} {\bibfnamefont {Mark J~H}\ \bibnamefont {Ku}},
  \bibinfo {author} {\bibfnamefont {Lawrence~W}\ \bibnamefont {Cheuk}},
  \bibinfo {author} {\bibfnamefont {Wenjie}\ \bibnamefont {Ji}}, \bibinfo
  {author} {\bibfnamefont {Waseem~S}\ \bibnamefont {Bakr}}, \ and\ \bibinfo
  {author} {\bibfnamefont {Martin~W}\ \bibnamefont {Zwierlein}},\ }\bibfield
  {title} {\enquote {\bibinfo {title} {{Heavy solitons in a fermionic
  superfluid}},}\ }\href {\doibase 10.1038/nature12338} {\bibfield  {journal}
  {\bibinfo  {journal} {Nature}\ }\textbf {\bibinfo {volume} {499}},\ \bibinfo
  {pages} {426--30} (\bibinfo {year} {2013})}\BibitemShut {NoStop}%
\bibitem [{\citenamefont {Paredes}\ \emph {et~al.}(2004)\citenamefont
  {Paredes}, \citenamefont {Widera}, \citenamefont {Murg}, \citenamefont
  {Mandel}, \citenamefont {F\"{o}lling}, \citenamefont {Cirac}, \citenamefont
  {Shlyapnikov}, \citenamefont {H\"{a}nsch},\ and\ \citenamefont
  {Bloch}}]{Paredes2004}%
  \BibitemOpen
  \bibfield  {author} {\bibinfo {author} {\bibfnamefont {Bel\'{e}n}\
  \bibnamefont {Paredes}}, \bibinfo {author} {\bibfnamefont {Artur}\
  \bibnamefont {Widera}}, \bibinfo {author} {\bibfnamefont {Valentin}\
  \bibnamefont {Murg}}, \bibinfo {author} {\bibfnamefont {Olaf}\ \bibnamefont
  {Mandel}}, \bibinfo {author} {\bibfnamefont {Simon}\ \bibnamefont
  {F\"{o}lling}}, \bibinfo {author} {\bibfnamefont {Ignacio}\ \bibnamefont
  {Cirac}}, \bibinfo {author} {\bibfnamefont {Gora~V}\ \bibnamefont
  {Shlyapnikov}}, \bibinfo {author} {\bibfnamefont {Theodor~W}\ \bibnamefont
  {H\"{a}nsch}}, \ and\ \bibinfo {author} {\bibfnamefont {Immanuel}\
  \bibnamefont {Bloch}},\ }\bibfield  {title} {\enquote {\bibinfo {title}
  {{Tonks-Girardeau gas of ultracold atoms in an optical lattice.}}}\ }\href
  {\doibase 10.1038/nature02530} {\bibfield  {journal} {\bibinfo  {journal}
  {Nature}\ }\textbf {\bibinfo {volume} {429}},\ \bibinfo {pages} {277--81}
  (\bibinfo {year} {2004})}\BibitemShut {NoStop}%
\bibitem [{\citenamefont {Kinoshita}\ \emph {et~al.}(2004)\citenamefont
  {Kinoshita}, \citenamefont {Wenger},\ and\ \citenamefont
  {Weiss}}]{Kinoshita2004}%
  \BibitemOpen
  \bibfield  {author} {\bibinfo {author} {\bibfnamefont {Toshiya}\ \bibnamefont
  {Kinoshita}}, \bibinfo {author} {\bibfnamefont {Trevor}\ \bibnamefont
  {Wenger}}, \ and\ \bibinfo {author} {\bibfnamefont {David~S}\ \bibnamefont
  {Weiss}},\ }\bibfield  {title} {\enquote {\bibinfo {title} {{Observation of a
  one-dimensional Tonks-Girardeau gas}},}\ }\href {\doibase
  10.1126/science.1100700} {\bibfield  {journal} {\bibinfo  {journal}
  {Science}\ }\textbf {\bibinfo {volume} {305}},\ \bibinfo {pages} {1125--8}
  (\bibinfo {year} {2004})}\BibitemShut {NoStop}%
\end{thebibliography}
\end{document}